\documentclass[Afour,sagev,times]{sagej}

\usepackage{moreverb,url}
\usepackage{bm} 
\usepackage{bbm} 
\usepackage{soul}
\usepackage{mathrsfs}
\usepackage{enumitem} 
\usepackage{varwidth} 
\usepackage{booktabs}

\usepackage{hyperref}
\hypersetup{colorlinks,bookmarksopen,bookmarksnumbered,citecolor=red,urlcolor=red}

\newcommand\BibTeX{{\rmfamily B\kern-.05em \textsc{i\kern-.025em b}\kern-.08em
T\kern-.1667em\lower.7ex\hbox{E}\kern-.125emX}}

\def\volumeyear{2021}
\newcommand{\nc}{\newcommand}
\nc{\yobs}{Y^\text{obs}}
\nc{\ymis}{Y^\text{mis}}
\nc{\xmis}{\bm{X}^\text{mis}}
\nc{\xobs}{\bm{X}^\text{obs}}
\nc{\zobs}{\bm{Z}^\text{obs}}
\nc{\zmis}{\bm{Z}^\text{mis}}
\nc{\zobsb}{\bm{Z}^{\text{obs}^{(b)}}}
\nc{\zimpb}{\bm{Z}^{\text{imp}^{(b)}}}
\nc{\cond}{{\, \vert \,}}
\nc{\indep}{{\, \perp \! \! \! \perp  \,} }
\nc{\tsps}{^{ {\rm T} } }
\nc{\code}[1]{\texttt{#1}}
\nc{\R}{{\normalfont\textsf{R}}{}}
\nc{\tb}{\textcolor{blue}}

\begin{document}

\runninghead{Hu et al.}

\title{Variable selection with missing data in both covariates and outcomes: Imputation and  machine learning}

\author{Liangyuan Hu\affilnum{1}, Jung-Yi Joyce Lin\affilnum{2}, Jiayi Ji\affilnum{2}}

\affiliation{\affilnum{1}Department of Biostatistics and Epidemiology,
Rutgers University School of Public Health, USA\\
\affilnum{2}Department of Population Health Science \& Policy, Icahn School of Medicine at Mount Sinai, USA
}

\corrauth{Liangyuan Hu, Department of Biostatistics and Epidemiology,
Rutgers University School of Public Health,
683 Hoes Lane West, Piscataway, NJ, USA 08854.}

\email{liangyuan.hu@rutgers.edu}

 \begin{abstract}
The missing data issue is ubiquitous in health studies. Variable selection in the presence of both missing covariates and outcomes is an important statistical research topic but has been less studied. Existing literature focuses on parametric regression techniques that provide direct parameter estimates of the regression model. In practice, parametric regression models are often  sub-optimal for variable selection because they are susceptible to misspecification. Flexible nonparametric machine learning methods considerably mitigate the reliance on the parametric assumptions, but do not provide as naturally defined variable importance measure as the covariate effect native to parametric models. We investigate a general variable selection approach  when both the covariates and outcomes can be missing at random and have general missing data patterns. This approach exploits the flexibility of machine learning modeling techniques and bootstrap imputation, which is amenable to nonparametric methods in which the covariate effects are not directly available. We conduct expansive simulations investigating the practical operating characteristics of the proposed variable selection approach, when combined with four tree-based machine learning methods, XGBoost, Random Forests, Bayesian Additive Regression Trees (BART) and Conditional Random Forests, and two commonly used parametric methods, lasso and backward stepwise selection. Numeric results suggest that  when combined with bootstrap imputation, XGBoost and BART have the overall best variable selection performance with respect to the $F_1$ score and Type I error, while the lasso and backward stepwise selection have subpar performance across various settings. In general, there is no significant difference in the variable selection performance due to imputation methods. Guidance for choosing methods appropriate to the structure of the analysis data at hand are discussed. We further demonstrate the methods via a case study of risk factors for 3-year incidence of metabolic syndrome with data from the Study of Women's Health Across the Nation.

\end{abstract}


\maketitle

\section{Introduction}
 Variable selection is an important statistical research topic concerning the identification of important predictors of an outcome variable.  The building block of variable selection is the modeling of the covariate-outcome relationship. Traditionally, parametric models have been used to describe how the covariates are related to the outcome; and variables can be selected based on hypothesis tests between nested models  or by optimizing a likelihood penalized for model complexity, typically using the Akaike or Bayesian information criterion. With parametric regression, the dependence structures among and distributional shapes of covariates need to be made explicit. Misspecification of these parametric forms will have an adverse effect on variable selection \citep{bleich2014variable}. Considerable research has demonstrated advantages of variable selection based on flexible machine learning modeling techniques compared to methods relying on parametric regression \citep{bleich2014variable, hu2020tree, hapfelmeier2013new, hu2020quantile, hu2020machine,ji2020identifying}. In particular, nonparametric tree-based methods including Bayesian Additive Regression Trees (BART) \citep{chipman2010bart}, Random Forests (RF) \citep{breiman2001random} and extreme gradient boosting (XGBoost) \citep{chen2016xgboost} have attracted a lot of attention in the scientific literature.  Tree-based machine learning methods mitigate the reliance on the parametric modeling assumptions and are more adept at capturing nonlinearities and interactions in the relationships among response and covariates \cite{breiman1984classification,chipman2010bart,hothorn2015partykit}, and have been widely applied  in health research \citep{hu2021estimation,hu2021estimating,hu2021estimatingh,hu2020estimation,hu2020tree,mazumdar2020comparison,hu2020ranking}.  Variable selection procedures for these nonparametric tree-based methods typically rely on the variable importance measure supplied by the model. For example, permutation-based \citep{bleich2014variable}  or sequential stepwise approaches \citep{diaz2006gene} have been used to select a set of important covariates that would optimize a performance metric, e.g., classification accuracy for a binary outcome or the mean-squared-error  for a continuous outcome. These procedures can better represent complex response surfaces but are ``black-boxes'' in the sense that the effects of covaraites on the outcome are not directly quantified \citep{bleich2014variable}.

 Missing data are ubiquitous in health care databases, and  present substantial challenges to variable selection, as variable selection procedures need to be tailored to various missing data mechanisms and statistical approaches used for handling missing data under the specific mechanisms. There are three general missing data mechanisms: missing completely at random (MCAR), missing at random (MAR), and missing not at random (MNAR) \citep{little2012prevention}. When the missingness depend neither on observed data nor on the missing data, the data are said to be MCAR. In this situation, complete cases analysis would  yield larger standard errors due to the reduced sample size but does not cause bias as the incomplete dataset is representative of the entire dataset \citep{sterne2009multiple}. More often the mechanism of missingness may depend on the observed data, and then the missing data are MAR given the observed data \citep{sterne2009multiple}. MAR allows prediction of the missing values based on the cases with complete data. In the situation where the missingness depends on the missing data, then the data are MNAR \citep{national2010prevention}.  Sensitivity analysis \citep{hogan2014bayesian,hu2018modeling} is an approach recommended by the National Research Council \citep{national2010prevention} to handle MNAR by assessing the impact of assumptions about the missing data on inference. In this paper, we focus on the mechanism of MAR, which is widely accepted in the biostatistical and health research, and imputation, a statistical technique for handing missing data that has gained wide popularity for its generality. Under MAR,  statistical techniques for treating missing data such as the inverse probability weighting \citep{tsiatis2007semiparametric} or the expectation-maximization algorithm \cite{garcia2010variable} have been used in combination of variable selection. However, these estimating equations-based or likelihood-based approaches are usually customized to a specific variable selection procedure and are less amenable to general missing data patterns \citep{long2015variable}.  Wood et al. \cite{wood2008should} compared strategies for combining variable selection procedures and imputation for incomplete data. They focused on the backward stepwise selection approach based on the parameter estimates and their standard errors adjusted for imputation. Long and Johnson \citep{long2015variable} proposed a general resampling approach that combines bootstrap imputation and stability selection \citep{meinshausen2010stability}, which is based on randomized lasso, and used simulations to demonstrate that this approach had better performance compared with several existing methods for both low- and high-dimensional problems. Both studies rely on parametric modeling of the covariate-outcome relationship.

Some tree-based  methods have embedded algorithms for handling  missing data. For example, BART and XGBoost can accommodate missing data in covariates by considering the missingness as a value in its own right in the tree splitting rules during model construction \citep{kapelner2015prediction,chen2016xgboost}. This technique is known as missingness incorporated in attributes (MIA). MIA does not require any assumptions or need for imputation, but cannot simultaneously address missing data in the outcome.  Some RF based algorithms do not distinguish between outcome and covariates and can deal with missing values in both cases and at the same time. Examples of such algorithms \citep{tang2017random} include 1) ``strawman imputation'', which imputes the missing values for continuous variables using the median of non-missing values and for discrete variables using the most frequently occurring non-missing values, 2) ``on-the-fly-imputation'',  which simultaneously imputes data by randomly drawing a value from in-bag data while growing the forest and iterating for improved results, and 3) imputation algorithm called missForest (more detail in Section~\ref{sec:vs-in-bi}), which recasts the missing data problem as a prediction problem. Using simulations, Tang et al.\citep{tang2017random} recommended the imputation method missForest, which performed the best across various missing data settings. To the best of our knowledge, no previous research has investigated strategies for combining statistical techniques for missing data and variable selection procedures based on flexible tree-based machine learning methods.


In this paper, we investigate approaches to combining imputation and variable selection procedures for tree-based methods. We consider and compare four machine learning methods that have a proven track record: BART, RF, conditional random forests (CRF) \citep{hothorn2006unbiased} (a variant of RF utilizing conditional inference trees as base learners), and XGBoost, and two commonly used classical parametric regression based methods: the backward stepwise selection approach and lasso. 
Drawing strength from previous work by Long and Johnson \citep{long2015variable}, we use bootstrap imputation \citep{efron1994missing}. It has been shown, when used with lasso, to be relatively insensitive to tuning parameter values in terms of variable selection, as well as the threshold value $\pi$ that determines which variables should be selected, e.g., if they are selected in at least $\pi M$ imputed datasets. Motivated by our case study, we focus on a binary outcome.  However, all methods considered can be straightforwardly applied to a continuous outcome.

The rest of the paper is organized as follows. Section~\ref{sec:methods} describes the methods.  In Section~\ref{sec:sim}, we conduct an expansive simulation to assess the practical operating characteristics of the methods. Section~\ref{sec:application} illustrates the methods using data from the Study of Women's Health Across the Nation, and Section~\ref{sec:discussion} provides a discussion.

\section{Methods}\label{sec:methods}
\subsection{Notation}
Suppose we have $n$ independently and identically distributed copies of data $\{(Y_i, \bm{X}_i), i =1,\ldots, n\}$, where $Y_i$ is subject $i$'s  binary outcome with 1 for case and 0 for control, $\bm{X}_i = (X_{i1}, \ldots, X_{iK})$ is a set of $K$ potential predictors, either continuous or discrete. Both $Y$ and $\bm{X}$ can have missing values. Let $R_k^X$ be the missing data indicator for $X_k$, with $R^X_k=1$ if $X_k$ is missing and $R^X_k=0$ if $X_k$ is observed, $k=1,\ldots,K$, and $\bm{R^X} = (R^X_1,\ldots, R^X_K)$. Similarly, let $R^Y$ be the missing data indicator for the outcome $Y$, with $R^Y =1$ if $Y$ is missing and $R^Y =0$ if $Y$ is observed.  Let $\bm{Q}$ denote the nonlinear transformations and higher-order terms of the predictors $\bm{X}$, and $\bm{R^Q}$ be the corresponding missing data indicator. We use $\bm{Z} = (\bm{X}, \bm{Q}, Y)$ to denote the full data,  $\zobs$ to indicate the observed components of $\bm{Z}$ and $\zmis$ the missing components. The missing data indicator for the full data is $\bm{R} = (\bm{R^X}, \bm{R^Q}, R^Y)$. Throughout, we assume that data are MAR, i.e., $f(\bm{R} \cond \bm{Z}) = f(\bm{R} \cond \bm{Z}^\text{obs})$.

\subsection{Overview of variable selection methods in complete data}
Variable selection has been widely studied from both the classical and Bayesian perspective. A large body of work rests on the assumption taking the shape of parametric models, in which the outcome is explicitly specified in a linear functional form of the covariates. If the parametric form is incorrectly specified, the variable selection results may be erroneous. For example, noise variables may be selected and important variables excluded. In practice, correctly specifying a parametric model in the presence of complex nonlinear functional relationships between the predictors and response is difficult. However, nonparametric methods are flexible to approximate functional forms of arbitrary complexity, thus mitigate the parametric assumptions and attendant errors in variable selection. In this article, we focus on tree-based
methods, examples of which include BART, RF, CRF and XGBoost.  The tree-based methods use the internals of the classification and regression tree (CART) structure. A CART model starts with a single root node containing all observed data. A splitting rule, involving a splitting variable $X_k$ and a split point $c$, is chosen for the root node. The data in the root node are then divided into two groups -- daughter nodes -- based on whether $X_k \geq c$ or $X_k <c$. This procedure is then sequentially performed on each of the daughter nodes until some stopping criteria have been met, leading down to the terminal nodes. See Breiman et al. \citep{breiman1984classification} for a detailed description of the CART. An ensemble of CART models can overcome the instability of single trees and improve prediction accuracy.  

\subsubsection{Bayesian Additive Regression Trees}
BART is a Bayesian sum-of-trees model  with a regularizing prior to
keep the individual tree effects small \citep{chipman2010bart}.  For a binary outcome, BART uses probit regression, 
\begin{eqnarray}
\mathrm{Pr}(Y= 1 \cond \bm{X}=\bm{x}) = \Phi\left(f(\bm{x})\right) = \Phi\left(\sum_{j=1}^m g(\bm{x}; \mathcal{T}_j, \mathcal{M}_j)\right),
\end{eqnarray}
where $\Phi (\cdot)$ is  the standard normal cumulative distribution function,  $\mathcal{T}_j$ denotes the $j$th regression tree with $v$ terminal nodes, each of which has a parameter $\mu_{jv}$ representing the mean response of the observations in the $d$th terminal node ($d=1, \ldots,v$), and $\mathcal{M}_j = (\mu_{j1}, \ldots, \mu_{jv})$;  and $g(\bm{x}; \mathcal{T}_j, \mathcal{M}_j)$ is the value obtained by dropping the vector of covariates $\bm{x}$ down the tree $\mathcal{T}_j$ and reporting the mean response $\mu$ associated with the terminal node in which $\bm{x}$ falls.  The $(\mathcal{T}_j, \mathcal{M}_j)$ are treated as parameters in a formal statistical model. To avoid overfitting and limit the contribution of each $(\mathcal{T}_j, \mathcal{M}_j)$, a regularizing prior is put on the parameters, and the posterior is computed using Markov chain Monte Carlo (MCMC). At each iteration of the MCMC algorithm, each tree may grow or shrink in size, with its parameters $(\mathcal{T}_j, \mathcal{M}_j)$ possibly swapped with another pair without changing the $f$. This lack of identification leads to a stable MCMC algorithm. A detailed overview of the BART model can be found in Chipman et al. \citep{chipman2010bart} and Bleich et al. \citep{bleich2014variable}.

Variable selection for BART draws on a permutation-based inferential approach \citep{bleich2014variable}.  The BART model outputs the ``variable inclusion proportions'' of each predictor variable: the proportion of times each predictor is chosen as a splitting rule divided by the total number of splitting rules appearing in the model. The variable inclusion proportions can be used to rank variables in terms of relative importance, but do not provide guidelines for variable selection. Bleich et al. \citep{bleich2014variable} proposed a nonparametric approach to establish thresholds for the variable inclusion proportions for a variable to be deemed as important. This approach is based on permutation of the response variable and averages out the chance capitalization that may occur in a single data set. Specifically, $P$ permutations of the response vector are created, $Y^*_1, \ldots, Y^*_P$. The BART model will then be fitted to each of the permuted response vectors $Y^*_p$ and the original predictor variables $(X_1, \ldots,X_K)$. From the BART run using each permuted response $Y^*_p$, we retain the variable inclusion proportions, $p^*_{k,p}$  for each predictor $X_k$. Denote the vector of all variable inclusion proportions from the $p$th permuted response by $\bm{\mathrm{p}}_p^*$, then the ``null'' distribution of the variable inclusion proportions across all $P$ permutations can be computed, $\bm{\mathrm{p}}_1^*, \ldots, \bm{\mathrm{p}}_P^*$, for the variable inclusion proportions $\bm{\mathrm{p}}$ from the real response $Y$. Three thresholding procedures of varying stringency are proposed to select important variables. The least stringent strategy is a ``local" threshold: select predictor $X_k$ if $p_k$ exceeds the $1-\alpha$ quantile of the permutation null distribution of $p_k$, $p_{k,1}^*, \ldots, p_{k,P}^*$. 
The most stringent strategy is a ``global max" threshold: first calculate $p^*_{\max, p} = \max\{p^*_{1,p}, \ldots, p^*_{K,p}\}$, the largest variable inclusion proportion across all predictor variables in permutation $p$, and then only select $X_k$ if $p_k$ exceeds the $1-\alpha$ quantile of the distribution of $p^*_{\max, 1}, \ldots, p^*_{\max, P}$. The intermediary strategy is a ``global SE" threshold: calculate the mean $m_k$ and standard deviation $s_k$  of variable inclusion proportion $p^*_k$ for predictor $X_k$ across all permutations, and find the smallest global multiplier $C^*$ such that $C^* = \inf\limits_{C\in {\mathbb{R}^+}} \left\{\forall k, \frac{1}{P} \sum_{p=1}^P \mathbbm{1} \left(p^*_{k,p} \leq m_k + C s_k \right) >1-\alpha\right\}$. The predictor $X_k$ is only selected if $p_k > m_k + C^* s_k$. 
In our simulations (Section~\ref{sec:sim}), we used the local threshold as the other two thresholds were found to be too stringent for our study settings. 
We  refer to Bleich et al. \citep{bleich2014variable} for more detail on the permutation-based variable selection approach. 

\subsubsection{Random Forests}
The RF method generates a distribution of trees,  each constructed on a bootstrap sample. This process is known as \emph{bagging}, short for bootstrap aggregation, first developed by Breiman as one of the earliest  ensemble techniques \citep{breiman1996bagging}. To reduce correlation among bootstrapped trees and improve the prediction accuracy of the ensemble model, Breiman \citep{breiman2001random} unified an algorithm called RF, which considers a random subset of predictors for each split in the tree-building process on each bootstrap sample. Certain observations of each bootstrap sample are left out and not used for fitting the tree model. These samples are called out-of-bag (OOB) samples and can be used to evaluate the predictive performance of the RF model.

RF supplies the OOB variable importance score, describing the relative impact of each predictor variable on the model's predictive accuracy. The score of each predictor variable is calculated by permuting that variable's values in the OOB sample while keeping all other variables the same, and recording the difference in prediction accuracy between the permuted and original data. Many variable selection procedures for RF are based on the combination of variable importance and model selection, in particular in the family of the ``wrapper'' methods. In this article, we use an approach that is based on recursive elimination of variables and has been widely used in the biomedical research. 

Following D{\'\i}az-Uriarte and De Andres \citep{diaz2006gene}, we iteratively fit a series of RF models. In each iteration, a new forest is built after a fraction of predictor variables are discarded with the smallest variable importance scores. The variable importance scores are not recalculated in each iteration to avoid potential overfitting. The OOB error rates from all the fitted RF are recorded. The selected set of variables is the smallest number of variables whose OOB error rate is within $u$ standard errors of the minimum error rate of all RF models (the standard error is computed from the exact binomial distribution with the rate parameter being the minimum error rate). Setting $u =0$ corresponds to selecting the set of variables that leads to the smallest error rate; and setting $u =1$ is similar to the ``1 s.e. rule" commonly used in the classification literature. We use $u=1$ in our simulations (Section~\ref{sec:sim}) and case study (Section \ref{sec:application}). 

\subsubsection{Conditional Random Forest}
CRF is a variant of the RF model with the conditional inference trees of Hothorn et al.\citep{hothorn2006unbiased} as base learners.
In a conditional inference framework, statistical hypothesis tests are utilized to conduct an exhaustive search across the predictor variables and the possible split points of these variables. For a candidate split, a statistical test with a $p$-value computed is used to evaluate the between-group difference for the two nodes generated by the split. A threshold for statistical significance is used as the stopping criterion to
determine whether additional splits should be performed. Like CART trees, conditional inference trees can  be bagged. In constructing the ensembles, observations are sampled according to probabilities specified by the weights. The CRF also differs from RF in  the aggregation scheme.  The CRF algorithm averages observation weights extracted from each of the conditional inference trees rather than direct predictions as  RF. For variable selection using CRF, we also employ the recursive backward elimination approach as used for RF, based on the variable importance score and OOB error rate outputted by CRF.  Hothorn et al. \citep{hothorn2006unbiased} and Strobl et al. \citep{strobl2007bias} argued that  computation of variable importance is biased in favor of variables with many potential cutpoints in RF; on the contrary, the conditional inference trees used in CRF are unbiased.

\subsubsection{Extreme Gradient Boosting}
The XGBoost method draws on the idea of gradient boosting \citep{chen2016xgboost}. Boosting is a process in which a weak learner  is boosted into a strong learner.  Friedman et al. \cite{friedman2000additive} connected boosting to a forward
stagewise additive model that minimizes a loss function  and brought forth a highly adaptable algorithm, gradient boosting machines. Specifically, a gradient tree boosting model uses $m$ additive functions to predict the outcome, 
\begin{eqnarray}\label{eq:gbm}
\mathrm{Pr} (Y=1 \cond \bm{X} = \bm{x}) = \phi(\bm{x})= \sum\limits_{j=1}^m f_j(\bm{x}),
\end{eqnarray} 
where $f = w_{q(\bm{x})}$ is a tree structure with decision rules $q$ that map an observation $\bm{x}$ to a terminal node, and $w$ represents the weights associated with terminal nodes.  Each tree contains a continuous score on each of the terminal nodes. For a given $\bm{x}$, we will drop $\bm{x}$ down each of the $m$ boosted trees until it hits a terminal node of each tree, and calculate the final prediction for $\bm{x}$ by summing up the scores (given by $w$) in the corresponding terminal nodes across the $m$ trees, i.e. $\sum_{j=1}^m f_j(\bm{x})$ in equation~\eqref{eq:gbm}. The idea of gradient boosting is to minimize the loss function  $\mathcal{L}(\phi) = \sum_{i} l(y_i, \hat{y}_i)$, where $\hat{y}_i$ and $y_i$ are respectively the predicted and observed outcome for the $i$th individual in the data. For classification,  $l(y_i,\hat{y}_i) = \exp{(-y_i \hat{y}_i)}$ is an exponential loss function (differentiable convex) that measures the difference between $\hat{y}_i$ and $y_i$, $\forall i \in  \{1, \ldots, n\}$. The XGBoost algorithm adds a term $\Omega(f_j)$ to the loss function $\mathcal{L}(\phi)$ of the traditional gradient tree boosting to penalize model complexity. The revised loss function for XGBoost is 
\begin{eqnarray}
\mathcal{L}(\phi) = \sum_{i=1}^n l(\hat{y}_i, y_i)+\sum\limits_{j=1}^m\Omega(f_j),
\end{eqnarray}
where $\Omega(f) = \gamma T + \frac{1}{2}\lambda \left \Vert w \right \Vert^2$, $T$ is the number of terminal nodes in a tree, $\Vert w\Vert$ represents the $\ell_2$ norm of terminal node weights, $\gamma$ and $\lambda$ are tuning parameters governing further partitioning of the predictor space. The penalty term $\Omega(f)$ helps to smooth the final learnt weights $w$ to avoid overfitting. XGBoost uses two additional techniques, shrinkage and column subsampling, to further prevent overfitting. Shrinkage scales newly added weights by a factor after each step of tree boosting, as in traditional gradient tree boosting. The column subsampling technique is borrowed from the RF algorithm, which selects a random subset of predictors for split in each step of tree boosting\citep{chen2016xgboost}.

To the best of our knowledge, there is no principled method to variable selection using XGBoost. We propose to leverage the variable importance score provided by XGBoost and apply the recursive feature elimination procedure used for RF to select important predictors with XGBoost. Since XGBoost does not have OOB samples, in our simulation study (Section~\ref{sec:sim}), we evaluate the model classification error on a 50\% hold-out set for $n=1000$ and using 5-fold cross-validation for $n=250$. 

\subsubsection{Least absolute shrinkage and selection operator} Shrinkage methods are frequently used for variable selection by regularizing the coefficient estimates. The lasso is among the most popular shrinkage methods \citep{tibshirani1996regression}. Variable selection via lasso for a binary outcome is by maximizing a penalized version of the log-likelihood of a logistic regression, 
\begin{eqnarray}\label{eq:lasso}
\max_{\beta_0,\bm{\beta}} \left\{\sum_{i=1}^n \left[y_i(\beta_0+\bm{\beta}^{\top} \bm{x}_i)  -\log \left(1+e^{\beta_0+\bm{\beta}^{\top} \bm{x}_i}\right) \right] - \lambda \sum_{k=1}^K \vert \beta_k\vert\right\},
\end{eqnarray}
where $\beta_0$ is the intercept term of the logistic regression model, and $\bm{\beta}$ corresponds to a vector of coefficients of $\bm{X}$. Criterion~\eqref{eq:lasso} is concave, and a solution of $\{\beta_0,\bm{\beta}\}$ can be found using nonlinear programming methods. The $\ell_1$ penalty in~\eqref{eq:lasso} has an effect of forcing some $\beta_k$'s to be exactly equal to zero when the tuning parameter $\lambda$ is sufficiently large. In this sense, the lasso selects the best subset of predictors and hence performs variable selection. 

\subsubsection{Backward Stepwise Selection} Classical variable selection is usually implemented through recursive procedures such as forward, backward, and stepwise selection. Both backward and forward selection methods have drawbacks, and stepwise selection provides a compromise. The backward stepwise selection
starts with a model with all potential predictors. The backward selection process removes from the model the predictor that has the least impact on the fit (e.g., at $\alpha$ significance level). The forward selection process checks whether removed variables should be added back into the model  (e.g., at $\alpha_2$ significance level, $\alpha_2 = \alpha (1-\epsilon$) for $\epsilon$ small). There are other stepwise selection procedures that use Z-score or the Akaike  information criterion for the inclusion and deletion processes \citep{friedman2001elements}. In this paper, we use the same procedures described in Wood et al. \citep{wood2008should}, which are based on statistical testing and take proper account of the multiple testing issues.

\subsection{Variable selection with missing data}
\subsubsection{Generating missing data} \label{sec:amputation}
To accurately mimic realistic missingness problems, we use the multivariate amputation approach described in Schouten et al. \citep{schouten2018generating} It has been shown that stepwise univariate amputation procedure -- generating missing values for one variable at a time -- may not appropriately control the missingness percentage and may lead to an inflated observed data skewness  when the missing data proportion increases\citep{schouten2018generating}. The multivariate amputation procedure generates any missing data scenario precisely as desired with respect to the missingness percentage and the missing data mechanism. In addition, the degree of skewness in the observed data is not sensitive to the number of amputated variables. These operating characteristics allow for valid and fair evaluations of  variable selection methods in the presence of missing data. 

The multivariate amputation procedure first randomly divides the complete data into a certain number (can be one) of subsets based on the assumed number of missing data patterns. The size of these subsets may vary. The use of subsets allows the specification of any missing data pattern for any subgroups while establishing the desired missingness percentage in the merged data. For example, in our case study (Section~\ref{sec:application}), participants with certain characteristics may be reluctant to answer questions related to their dietary habits, which will lead to joint missingness in all dietary variables (e.g., dietary intakes of isoflavones, lignans or coumestrol) \citep{greendale2012dietary}. This missingness characteristic  can be emulated by generating joint missingness in several variables in a subset of data. The weighted sum scores are then calculated for individuals in each subset to amputate the data. Specifically, the weighted sum score for an MAR predictor $X_k$ and individual $i$ of a subset follows 
\begin{eqnarray} \label{eq:wss}
wss_{x_k,i} =  \bm{X}_i w_1 +  \bm{Q}_i w_2 +  Y_i w_3, \; w_1, w_2, w_3 \in \mathbb{R}
\end{eqnarray}
where $w_1, w_2$ and $w_3$ are respectively a vector of user-specified weights for the untransformed versions of the predictors  $\bm{X}_i$, transformed versions of the predictors captured in $\bm{Q}_i$ and the outcome $Y_i$. Note that the weights are the same for individuals in the same subset (assumed to have the same missing data pattern). To induce MAR, a zero weight is assigned to the predictor that will be amputated. For example, if $R^X_k=1$, then the weights for  $X_k$ and its nonlinear transformation and higher order terms will be set to zero;  while nonzero weights will be assigned to the other predictors and their nonlinear counterparts $\bm{X}_{-k}^s \subseteq (\bm{X}_{-k}, \bm{Q}_{-k})$, on which $p(R^X_k \cond \bm{X}_{-k}^s)$ depends. A logistic distribution function, $\mathrm{Pr}(R_k^X = 1)= \text{logit}^{-1}\left(g(\bm{X}_{-k}^s)\right)$ is then applied on $wss_i$ to compute the missingness probability, which is used to determine whether the data point becomes missing or not.   Different $g(\cdot)$ functions can be used to create different missing data patterns.  For example, the standard logistic function $g(x) = 1/(1+e^{-x})$ generates a right-tailed type of missingness, that is, individuals with high weighted sum scores will receive a high probability of being missing; $g(x) = -a + |x-b| \; \forall a \in \mathbb{R}^+, b\in \mathbb{R}$ creates a both-tailed missingness type giving higher missingness probabilities to individuals with extreme weighted sum scores \citep{van2018flexible}. Putting all the subsets together, we will have a complete data set with desired missingness characteristics. 

\subsubsection{Variable selection in bootstrap imputed data} \label{sec:vs-in-bi}
Following Long and Johnson\cite{long2015variable}, we first conduct bootstrap imputation -- originally developed by Efron\cite{efron1994missing} -- before variable selection.
Wood et al.\citep{wood2008should} investigated methods for combining multiple imputation and the backward stepwise  variable selection. Their recommended approaches relied on combined inference about parameter estimates to consolidate variable selection results.
These approaches, however, are not amenable to machine learning modeling techniques as the parametric functional forms are not required in machine learning models.  Another popular approach is to perform variable selection on each of the $M$ imputed datasets, and then select  predictors if they appear in at least $\pi M$ models. Results in Wood et al.\citep{wood2008should} show that this approach, when combined with multiple imputation, can be sensitive to the threshold value $\pi$. The numerical studies in Long and Johnson \citep{long2015variable} suggest that when combined with bootstrap imputation and stability selection, this approach can be relatively insensitive to $\pi$ and tuning parameter values used in variable selection. Stability selection is a general variable selection approach based on resampling or subsampling for fully observed data \citep{meinshausen2010stability}. In stability selection, the randomized lasso is applied to each random sample (by random subsampling or resampling) of the observed data, and important predictors are selected based on the variable selection results from all random samples using a threshold $\pi$. We investigate the operating characteristics of bootstrap imputation used in conjunction with machine learning based variable selection procedures. The bootstrap imputation procedure is described as follows: 
\begin{enumerate}
\item Generate $B$ bootstrap datasets $\{\bm{Z}^{(b)}, b=1,\ldots, B \}$, each of which has a missing indicator $\bm{R}^{(b)}$ corresponding to the observed data $\zobsb$. 
\item Conduct imputation for each bootstrap data set $\{\bm{Z}^{(b)}, \bm{R}^{(b)}\}$ using an imputation method of choice. The imputed and complete datasets are denoted by $\{\zimpb = (Y^{(b)}, \bm{X}^{(b)}), b = 1, \ldots, B\}$.
\end{enumerate}

A single imputation is performed for each bootstrap sample. We use two methods for imputation in Step 2: (i) the standard imputation program using the $\R$ package $\code{mice}$ \citep{buuren2010mice}, and (ii) RF based imputation algorithm using the $\R$ package $\code{missForest}$ \citep{stekhoven2012missforest}. Imputation via $\code{mice}$ uses the iterated chained equations approach \citep{van1999multiple}, which requires specifying conditional models for each incomplete variable given all other variables and drawing imputations by iterating over the conditional distributions. The $\code{missForest}$ approach imputes data by growing a forest and regressing  each variable that has missing values in turn against all other variables and then predicting missing data for the dependent variable using the fitted forest \citep{tang2017random}. The $\code{missForest}$ method has been shown to have a better imputation performance than the parametric $\code{mice}$ approach \citep{waljee2013comparison}. We investigate whether imputation via nonparametric $\code{missForest}$ can improve variable selection with missing data over parametric $\code{mice}$. 
Variable selection via each of the methods considered will be performed for all $b=1, \ldots, B$ bootstrap imputed datasets. Denote by $\hat{\mathcal{S}}^{(b)}$ the selected set of predictors on the $b$th bootstrap imputed data. Then the final selected set of predictors $\hat{\mathcal{S}}_\pi \subseteq \bm{X}$ is 
\begin{eqnarray}
\hat{\mathcal{S}}_\pi = \left\{X_k: \Pi_k \geq \pi \right\},
\end{eqnarray}
where  $\Pi_k = (1/B) \sum_{b=1}^B \mathbbm{1} \left(X_k \in \hat{\mathcal{S}}^{(b)} \right)$ and $\pi \in (0,1)$ is a fraction threshold for selecting a predictor. Higher values of $\pi$ correspond to more stringent rules for selecting the important variables, whereas lower values of $\pi$ can lead to larger sets of selected variables, which may include noise predictors. In our simulations (Section~\ref{sec:sim}), We explore the performance of all methods considered using the threshold values from the fine resolution grid on the interval $[0,1]$. 
\subsection{Performance metrics}
Each of the methods considered will be compared on the ability to select
 predictor variables that are truly associated with the outcome, or useful predictor variables. We use the following five performance metrics commonly used in the variable selection literature \cite{bleich2014variable, wood2008should}. 

 \begin{enumerate}[label={(\arabic*)}]
     \item  $\text{Precision} =\dfrac{\text{TP}}{\text{TP} + \text{FP}}$, where TP and FP are respectively  the number of true positive and false positive selections.  The \emph{precision} of a variable selection method is the proportion of truly useful predictors among all selected predictors. 
     \item  $\text{Recall} = \dfrac{\text{TP}}{\text{TP}+\text{FN}}$, where FN is the number of false negative selections. The \emph{recall} of a variable selection method is the proportion of truly useful variables selected among all useful variables. This is sometimes referred to as the \emph{power} in the literature \cite{wood2008should}.
     \item $F_1 =  \dfrac{2\text{ Precision} \cdot \text{Recall} }{\text{Precision} + \text{Recall}}.$ The $F_1$ measure is the harmonic mean of precision and recall, balancing a method's ability to avoid selecting irrelevant predictors (precision) with its ability to identify the full set of useful predictors (recall). 
     \item Type I error. A variable selection method's \emph{Type I error} is defined as the mean of the probabilities that the method will incorrectly select each of the noise predictors. 
 \end{enumerate}
A good method should have high values of precision, recall and $F_1$ , and low Type I error (which may differ from the nominal significance level used in some model selection procedures) \citep{wood2008should}. 
 
\subsection{Software implementation}
The multivariate amputation procedure was conducted using the $\code{ampute}$ function in the $\R$ package $\code{mice}$ \citep{schouten2018generating}. Parametric imputation was performed using the  $\R$ package $\code{mice}$ \citep{buuren2010mice}, and nonparametric imputation was performed using the  $\R$ package $\code{missForest}$  \citep{stekhoven2012missforest}. We used the $\R$ package $\code{bartMachine}$ to perform variable selection for BART. For the BART models, we used 1100 draws with the first 100 discarded as burn-in. The permutation distribution was obtained from 100 BART model constructions. To implement variable selection using RF, the $\R$ package $\code{varSelRF}$ was used\citep{diaz2006gene}. We used the $\code{cforest}$ function in the $\R$ package $\code{partykit}$\citep{hothorn2015partykit} to fit the CRF models. To build RF and CRF models for recursive elimination of predictors, we used 5000 trees for the first full model and 2000 trees in each iteration step. The fraction of variables from those in the previous forest to exclude at each iteration  was set to $\code{vars.drop.frac} = 0.1$, $\code{mtry}$ (the number of randomly selected variables) was set at $\sqrt{K}$ and default values for other tuning parameters (e.g., $\code{nodesize}$ =1 and $\code{maxnodes}$ was not specified to allow trees to be grown to the maximum possible for RF;  $\code{minbucket}$ = 7, $\code{minsplit}$ = 20 for CRF) were used.   The XGBoost models were fitted using the $\code{xgboost}$ function in the $\R$ package $\code{xgboost}$\citep{chen2021xgboost}; the max number of tree boosting iterations was set to $\code{nrounds} = 200$ and the fraction of variables to be excluded in each iteration was set as 0.1. We implemented the lasso  using the $\R$ package $\code{glmnet}$ \citep{friedman2010regularization}. The tuning parameter $\code{lambda}$ was selected  as the largest value such that the mean cross-validated deviance was within one standard error of the minimum deviance. $\R$ codes to implement stepwise selection and our proposed variable selection algorithms using XGBoost and CRF, and to replicate our simulation studies are provided in the GitHub page of the first author \url{https://github.com/liangyuanhu/Variable-selection-w-missing-data}. 
\section{Simulation} \label{sec:sim}
\subsection{Simulation design} \label{sec:simdesign}
The simulation scenarios are motivated by our case study to represent data structures commonly observed in health studies. 
We considered two sample sizes,  small sample size $n=250$ and large sample size $n=1000$. We generated 10 useful predictors that are truly related to the responses, $X_1, \ldots, X_{10}$, and 40 noise predictors, $X_{11}, \ldots, X_{50}$. This ratio of useful versus noise predictors was chosen to mimic our case study. In Section~\ref{sec:simresults}, we 
vary the number of irrelevant predictors from 10 to 100 and examined how different ratios will impact the performance of the methods considered.  We drew independently $X_1$ and $X_2$ from  $\text{Bern}(0.5)$, $X_3$, $X_4$ and $X_5$ from the standard $N(0,1)$, $X_6$ from $\text{Gamma}(4, 6)$, and $X_7, X_8, X_9, X_{10}$ were designed to have missing values under the MAR mechanism. For an MAR predictor, we specify the true forms in which  the predictor depends on the other predictors. Both nonlinearity and nonadditivity were considered for the dependence structure among the predictor variables to compare the imputation methods $\code{missForest}$ and $\code{mice}$.  Specifically, the following are the true data models for $X_7, X_8, X_9, X_{10}$:
\begin{eqnarray*}
&x_7 \cond x_5,x_6 \sim N(-0.4x_5+0.4x_6+0.3x_5 x_6,1)\\
&x_8 \cond x_5,x_6,x_7 \sim N(0.1x_5 (x_6-2)^2-0.1x_7^2,1)\\
&x_9 \cond x_3,x_4,x_5 \sim N(0.5x_3+0.3x_4-0.3x_5^2+0.2x_3 x_4,1)\\
&x_{10} \cond x_3,x_4,x_5,x_9  \sim N(0.1x_3^3-0.3x_4-0.4x_5+0.2x_9^2+0.3x_4 x_5,1).
\end{eqnarray*}
We further generated 20 continuous noise predictors $x_{11}, \ldots, x_{30} \stackrel{i.i.d}{\sim}N(0,1)$,  and 20 binary noise predictors $x_{31}, \ldots, x_{50} \stackrel{i.i.d}{\sim}\text{Bern}(0.5)$. The true outcome model is specified as the following: 
\begin{equation*}
\begin{split}
\mathrm{Pr}(y=1 \cond x_1,\ldots,x_{10}) = \text{logit}^{-1} \big(&-2.7+1.8x_1+0.5x_2+1.1x_3-0.4e^{x_5 }-0.4(x_6-3.5)^2+0.3(x_7-1)^3+1.1x_8 \\
&-1.1x_{10}+5\sin(0.1\pi x_4x_9)-0.4x_5 x_{10}^2+0.4 x_3^2 x_8\big).
\end{split}
\end{equation*}
To avoid any concerns that we are deliberately selecting simulation settings where machine learning methods will show good performance, we considered the outcome model with arbitrary data complexity that reflects common situations in health datasets: (i) discrete predictors with strong ($x_1$) and moderate ($x_2$) associations; (ii) both linear and nonlinear forms of continuous predictors; (iii) nonadditive effects ($x_4x_9$).  

After generating the full data, we amputated  $X_7, X_8, X_9, X_{10}$ and $Y$ under MAR using the multivariate amputation approach described in Section~\ref{sec:amputation}. Previous  studies \citep{madley2019proportion,hughes2019accounting} show that multiple imputation could provide unbiased results when the proportion of missing data is up to 90\%. We investigate the performance of bootstrap imputation based variable selection methods across three levels of missingness: (i)  40\% missingness in $Y$ and 60\% overall missingness; (ii) 15\% missingness in $Y$ and 30\% overall missingness; and (iii) 7.5\% missingness in $Y$ and 15\% overall missingness.  Scenario (i) mimics the percentages of missingness observed in our case study (Section~\ref{sec:application}). Note that each variable can still have a small-to-moderate proportion of missing values. The percentage of the cases that are made incomplete indicates the overall missingness. For scenario (i), we first randomly divided the full data into 8 subsamples, which were the following percentages of the whole data: 0.30,  0.09,  0.09, 0.08,  0.08, 0.16,  0.10, and 0.10. Each subsample was then amputated  to have a specific missing pattern using the weighted sum scores defined in equation~\eqref{eq:wss}. The weighted sum scores relate the missingness on amputated variables to the values of other variables  as follows: 
\begin{enumerate}[label={(\arabic*)}]
    \item $wss_{y,i}=5x_1+5x_2+x_3-x_5-x_6+x_7+x_8+x_{10}-0.5x_6^2+1.5x_4x_9-0.5x_5 x_{10}+0.5x_3 x_8$ 
    \item $wss_{x_7,i}=x_5+x_6+x_5 x_6$
    \item $wss_{x_8,i}=5y+x_5+x_6+x_7+x_7^2+x_5 x_6$
    \item $wss_{x_9,i}=5y+x_3+x_4+x_5+x_5^2+x_3 x_4$
    \item $wss_{x_{10},i}=5y+x_3+x_4+x_5+x_9+x_4 x_5$
    \item $wss_{y,x_7, x_8,i}=x_5+x_6$
     \item $wss_{y,x_8, x_{10},i}=x_5$
    \item $wss_{y,x_9, x_{10},i}=x_3+x_4+0.5x_5^2+0.5x_3 x_4$
\end{enumerate}
The weighted sum score gives a nonzero weight to the variables and their nonlinear forms and interactions therein, on which the probabilities to be missing for amputated variables depend.  In subsample (1), the responses were amputated.  The predictor variables $X_7, X_8, X_9, X_{10}$ were respectively amputated in subsample (2), subsample (3), subsample (4), and subsample (5). We further created the joint missingness in $(Y, X_7, X_8)$ in subsample (6), $(Y, X_8, X_{10})$ in subsample (7), and $(Y, X_9, X_{10})$ in subsample (8). Finally, we applied the logistic distribution function to the weighted sum scores to create the missing indicators and amputate data. A right-tailed type of missingness was used for subsamples (1)--(5) and a both-tailed type of missingness was used for subsamples (6)--(8). In each subsample,  there were 60\% missing values in the amputated variables. The amputated eight subsamples were combined to form a whole dataset. This setup creates 40\% missingness in $Y$ and 60\% overall missingness. The  scenario (iii) with 7.5\%  missingness in $Y$ and 15\% overall missingness was generated by creating 15\% missing values in the amputated variables. We generated  missingness scenario (ii) with 15\%  missingness in $Y$ and 30\% overall missingness by changing the proportions of the eight subsamples to 0.20, 0.15, 0.15, 0.10, 0.10, 0.10, 0.10, and 0.10 and creating 30\% missing values in each subsample.

There are a total of 8 scenarios considered in our simulation: 2 sample sizes ($n=250$ and $n=1000$) $\times$ 4 percentages of missingness (complete data, 15\% overall missingness, 30\% overall missingness and 60\% overall missingness). For computational considerations, we replicated each  simulation independently $M=250$ times for $n = 1000$ and $M=1000$ times for $n = 250$. For each Monte Carlo dataset, we draw $B=100$ bootstrap samples and perform imputation on each sample. In Section~\ref{sec:simresults}, we additionally investigated the sensitivity of complete data performance -- a benchmark against which performance on incomplete data will be compared -- to the number of noise predictor variables.  



\subsection{Simulation results}\label{sec:simresults}
When implementing the methods, we included all 50 predictor variables (10 useful and 40 noise predictors) and the outcome variable available to the analyst in the imputation models for both $\code{mice}$ and $\code{missForest}$. The top panel of Table~\ref{tab:simres-mice} displays  four performance metrics for each of six methods considered on complete data with  $n=250$ and $n=1000$. For the large sample size $n=1000$, four machine learning methods, BART, XGBoost, CRF and RF outperformed the two parametric methods, lasso and backward stepwise selection.  BART, XGBoost, CRF and RF all had good performance, with BART being the top performer and CRF and RF having relatively lower recall.   Between the two parametric methods, lasso tended to have higher precision and backward stepwise selection tended to higher recall and higher type I error.  A perusal of the recall for each of 10 useful predictor variables (Web Figure 1) shows that lasso and backward stepwise selection failed to identify the two nonadditive variables $X_4$ and $X_9$, and had a low probability of selecting $X_{10}$, which had complex forms of relationship with other predictors and with the response. On the other hand, BART and XGBoost had a high success rate of identifying all continuous variables in conditions of nonadditivity and nonlinearity, followed by CRF and RF, but all four machine learning based methods were less likely to select discrete variables compared to parametric methods. 
All methods had  deteriorated performance, with no apparent ``winning'' method when the sample size decreased from $n=1000$ to $n=250$, demonstrated by lower precision, recall and $F_1$ and higher Type I error.  The recall or power, in particular, had a substantial drop, suggesting the difficulty in recovering the full set of useful predictors with a small sample size. 

\setlength{\tabcolsep}{3pt} 
\begin{table}[htbp]
    \centering
    \caption{Simulation results for a combination of 6 scenarios = 2 samples sizes ($n=250$ and $n=1000$) $\times$ 4 percentages of missingness (complete data, 15\% overall missingness, 30\% overall missingness and 60\% overall missingness). There are a total of 50 predictors, 10 of which are truly related to the response variable and 40 are noise predictors. Imputation was conducted via $\code{mice}$. For each of six methods, we show results corresponding to the best threshold values of $\pi$ (based on $F_1$) as well as results on the complete cases (CC). Methods are sorted in descending order of $F_1$ for $n=1000$.} 
    \bgroup
\def\arraystretch{1.1} 
    \begin{tabular}{lccccclcccc}
    \toprule
         & \multicolumn{4}{c}{$n=250$} &&& \multicolumn{4}{c}{$n=1000$} \\\cline{2-5}\cline{8-11}
         &Precision &Recall &$F_1$  &Type I error &&&Precision &Recall &$F_1$  &Type I error\\\hline 
      \multicolumn{11}{c}{\textbf{Complete data}}\\
         BART& 0.87&0.44&0.58&0.02&&BART&1.00&0.87&0.93&0.00\\
        XGBoost& 0.82&0.53&0.62&0.04&&XGBoost&0.93&0.81&0.86&0.02\\
        CRF& 0.95&0.42&0.55&0.01&&CRF&1.00&0.72&0.83&0.00\\
         RF& 0.94&0.40&0.54&0.01&&RF&1.00&0.70&0.82&0.00\\
        Stepwise& 0.68&0.56&0.60&0.08&&Stepwise&0.79&0.74&0.76&0.05\\
         lasso& 0.83&0.36&0.51&0.03&&lasso&0.87&0.65&0.74&0.03\\
         \midrule
    \multicolumn{11}{c}{\textbf{7.5\% missingness in $Y$ and 15\% overall missingness}}\\
     BART $\pi=0.2$ & 0.67&0.58&0.64&0.04&&BART $\pi=0.1$ &0.99&0.92&0.96&0.00\\
      XGBoost $\pi=0.3$& 0.78&0.67&0.72&0.04&&XGBoost $\pi=0.4$&0.99&0.88&0.95&0.00\\
       CRF $\pi=0.3$ & 0.83&0.52&0.62&0.02&& CRF $\pi=0.2$ &0.96&0.90&0.92&0.02\\
       RF $\pi=0.1$ & 0.85&0.54&0.64&0.02&&RF $\pi=0.2$ &0.95&0.89&0.91&0.01\\
       Stepwise $\pi=0.5$ & 0.66&0.59&0.62&0.09&&Stepwise $\pi=0.6$ & 0.95&0.72&0.82&0.01\\
      lasso $\pi=0.7$ & 0.80&0.49&0.59&0.03&&lasso $\pi=0.8$&0.94&0.72&0.81 &0.01\\   
        
  \midrule
    BART CC& 0.62&0.51&0.56&0.10&&BART CC&0.91&0.85&0.88&0.08\\
 XGBoost CC & 0.67&0.59&0.61&0.06&&XGBoost CC&0.88&0.82&0.85&0.04\\     
 CRF CC& 0.50&0.55&0.53&0.08&&CRF CC&0.86&0.79&0.82&0.04\\
     RF CC & 0.50&0.54&0.52&0.09&&RF CC &0.85&0.78&0.81&0.03\\
     Stepwise CC&0.82&0.37&0.50 &0.03&&Stepwise CC&0.86&0.67&0.75&0.03\\
       lasso CC& 0.91&0.30&0.45&0.03&& lasso CC&0.85&0.66&0.73&0.03\\
       \midrule
    \multicolumn{11}{c}{\textbf{15\% missingness in $Y$ and 30\% overall missingness}}\\
         BART $\pi=0.2$ & 0.64&0.55&0.59&0.08&&BART $\pi=0.1$&0.97&0.89&0.92&0.01\\
        XGBoost $\pi=0.3$& 0.77&0.59&0.66&0.05&&XGBoost $\pi=0.4$& 0.97&0.83&0.89&0.01\\
         RF $\pi=0.2$ & 0.76&0.52&0.62&0.04&& RF $\pi=0.1$&0.91&0.84&0.87&0.02\\
         CRF $\pi=0.2$ & 0.77&0.53&0.63&0.03&& CRF $\pi=0.2$ &0.98&0.77&0.86&0.00\\
        Stepwise $\pi=0.5$ & 0.60&0.55&0.56&0.11&&  Stepwise $\pi=0.7$ &0.89&0.69&0.77&0.02\\
         lasso $\pi=0.7$ & 0.73&0.43&0.52&0.05&&lasso $\pi=0.8$&0.90&0.64&0.74&0.02\\
         \midrule
          BART CC  & 0.56&0.47&0.51&0.10&&BART CC &0.93&0.79&0.85&0.02\\
      XGBoost CC  & 0.82&0.44&0.56&0.05&& XGBoost CC &0.87&0.77&0.81&0.04\\  
         CRF CC & 0.44&0.56&0.48&0.13&&CRF CC & 0.82&0.76&0.79&0.03\\      
            RF CC & 0.43&0.55&0.47&0.15&&RF CC &0.84&0.75&0.78&0.04\\
      Stepwise CC &0.78&0.29&0.42 &0.02&& Stepwise CC &0.80&0.62&0.71&0.04\\
         lasso CC & 0.88&0.23&0.36&0.03&& lasso CC &0.80&0.60&0.70&0.04\\
         \midrule
     \multicolumn{11}{c}{\textbf{40\% missingness in $Y$ and 60\% overall missingness}}\\
    XGBoost $\pi=0.3$ & 0.55&0.60&0.57&0.13&&XGBoost $\pi=0.5$&0.95&0.73&0.82&0.01\\
 BART $\pi=0.2$ & 0.47&0.47&0.47&0.14&&BART $\pi=0.2$&0.97&0.66&0.78&0.01\\
  CRF $\pi=0.2$ & 0.55&0.52&0.52&0.11&&CRF $\pi=0.2$&0.91&0.69&0.78&0.02\\
         RF $\pi=0.2$ & 0.53&0.51&0.51&0.12&&RF $\pi=0.2$&0.91&0.67&0.77&0.02\\
 Stepwise $\pi=0.5$ & 0.48&0.48&0.46&0.15&&  Stepwise $\pi=0.8$&0.93&0.62&0.74&0.01\\
         lasso $\pi=0.8$ & 0.56&0.35&0.42&0.08&& lasso $\pi=0.9$&0.93&0.59&0.71&0.01\\
         \midrule
          XGBoost CC & 0.56&0.52&0.51&0.07&& XGBoost CC &0.90&0.68&0.77&0.04\\
          BART CC & 0.43&0.45&0.44&0.15&&   BART CC &0.91&0.60&0.75&0.03\\
        CRF CC & 0.46&0.43&0.45&0.14&&CRF CC &0.86&0.62&0.72&0.04\\
         RF CC & 0.44&0.43&0.44&0.15&&RF CC&0.85&0.60&0.71&0.05\\
           Stepwise CC &0.72&0.25&0.37 &0.10&& Stepwise CC &0.87&0.57&0.69&0.04\\
         lasso CC & 0.80&0.21&0.33 &0.06&& lasso CC &0.87&0.53&0.66&0.03\\
    \bottomrule
    \end{tabular}
  \egroup  
    \label{tab:simres-mice}
\end{table}

\begin{figure}[!ht]
    \centering
    \includegraphics[width=1\textwidth]{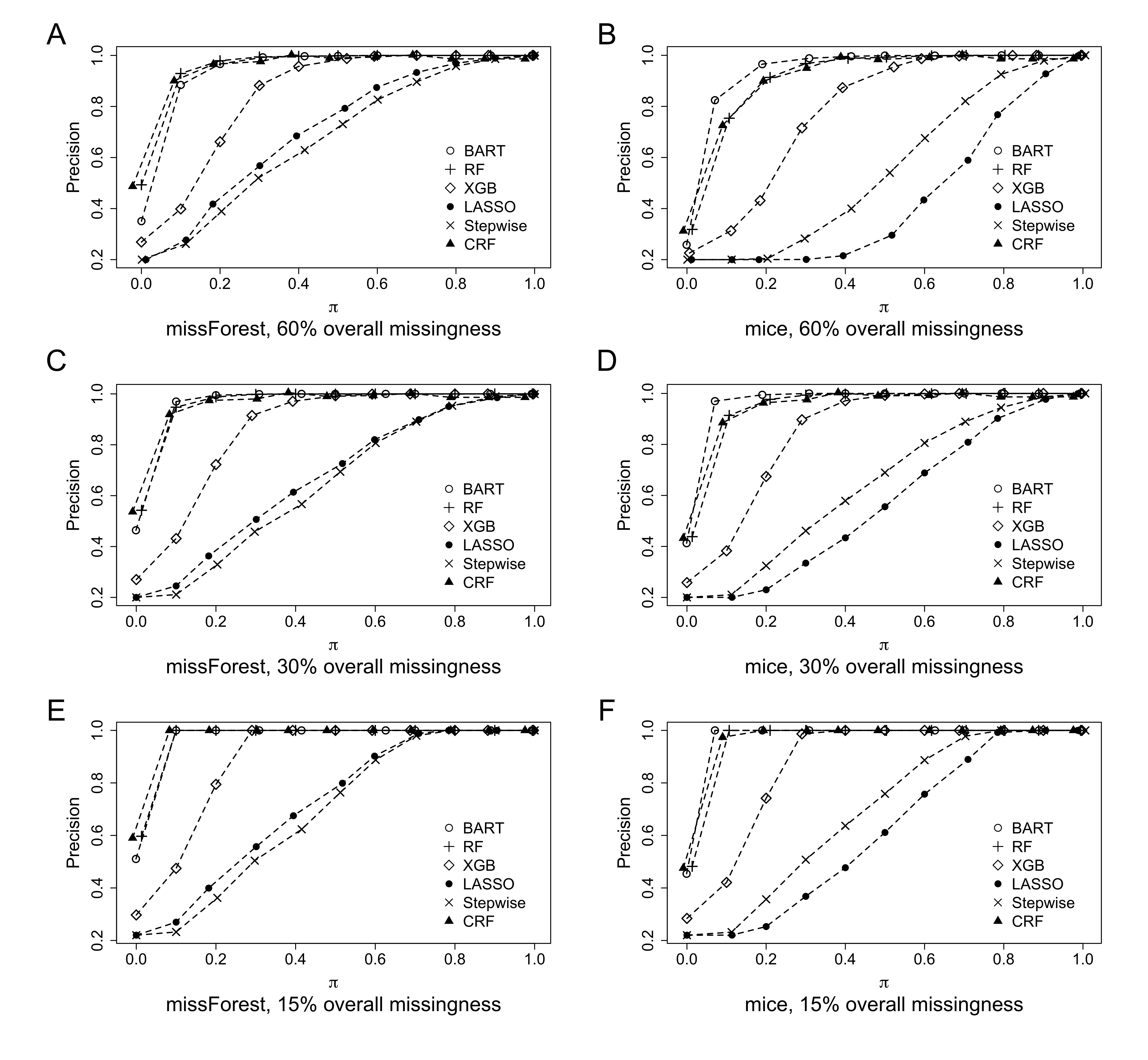}
    \caption{The precision by threshold values of $\pi$ for each of six variable selection methods and for sample size $n=1000$, based on 250 replications.  Imputation was performed on 100 bootstrap samples of each replication data set, using two imputation methods $\code{mice}$ and $\code{missForest}$ for each of three overall missingness proportions, 15\%, 30\% and 60\%.  } 
    \label{fig:precision_n_1000}
\end{figure}

\begin{figure}[!ht]
    \centering
    \includegraphics[width=1\textwidth]{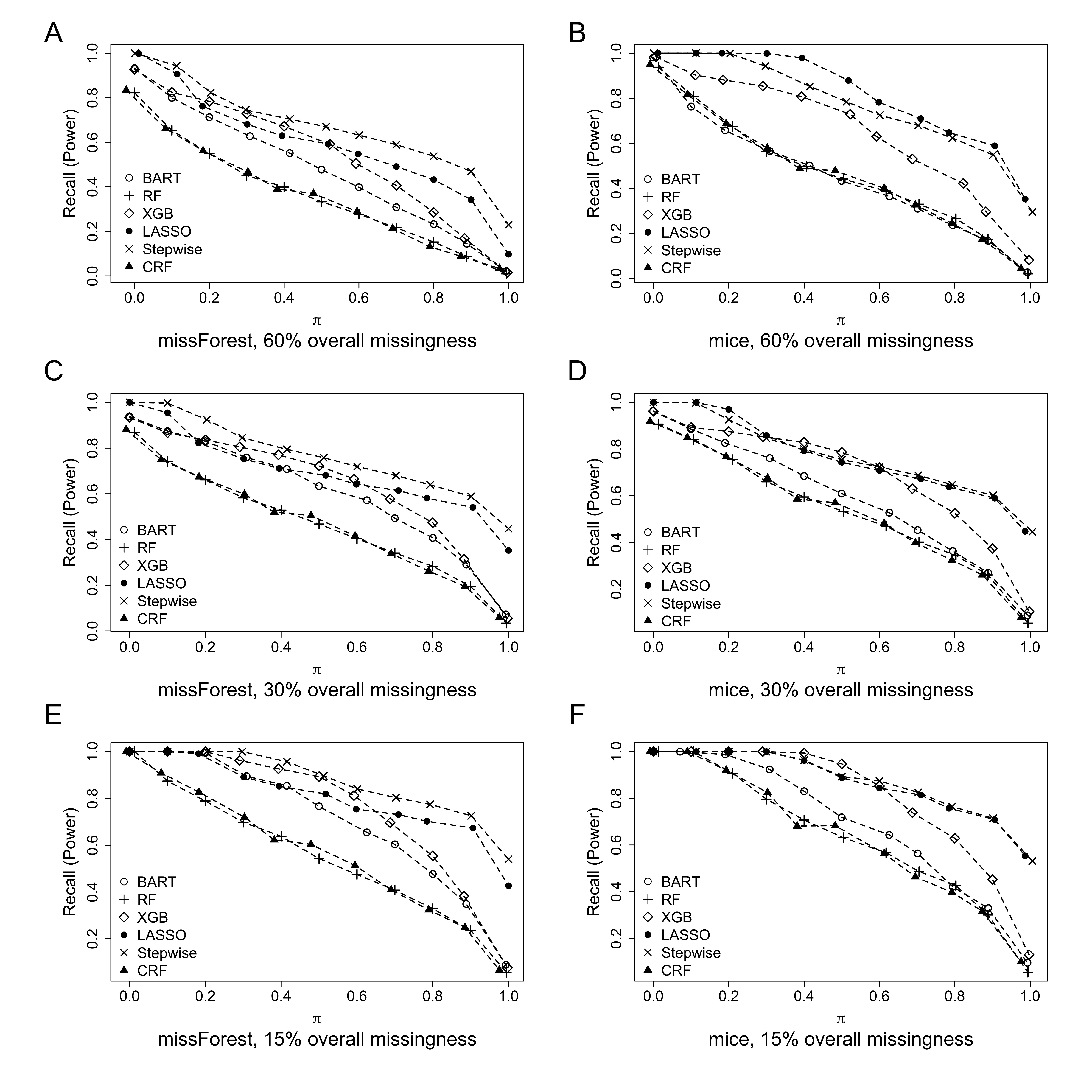}
    \caption{The recall by threshold values of $\pi$ for each of six variable selection methods and for sample size $n=1000$, based on 250 replications.  Imputation was performed on 100 bootstrap samples of each replication data set, using two imputation methods $\code{mice}$ and $\code{missForest}$ for each of three overall missingness proportions, 15\%, 30\% and 60\%. } 
    \label{fig:recall_n_1000}
\end{figure}

\begin{figure}[!ht]
    \centering
    \includegraphics[width=1\textwidth]{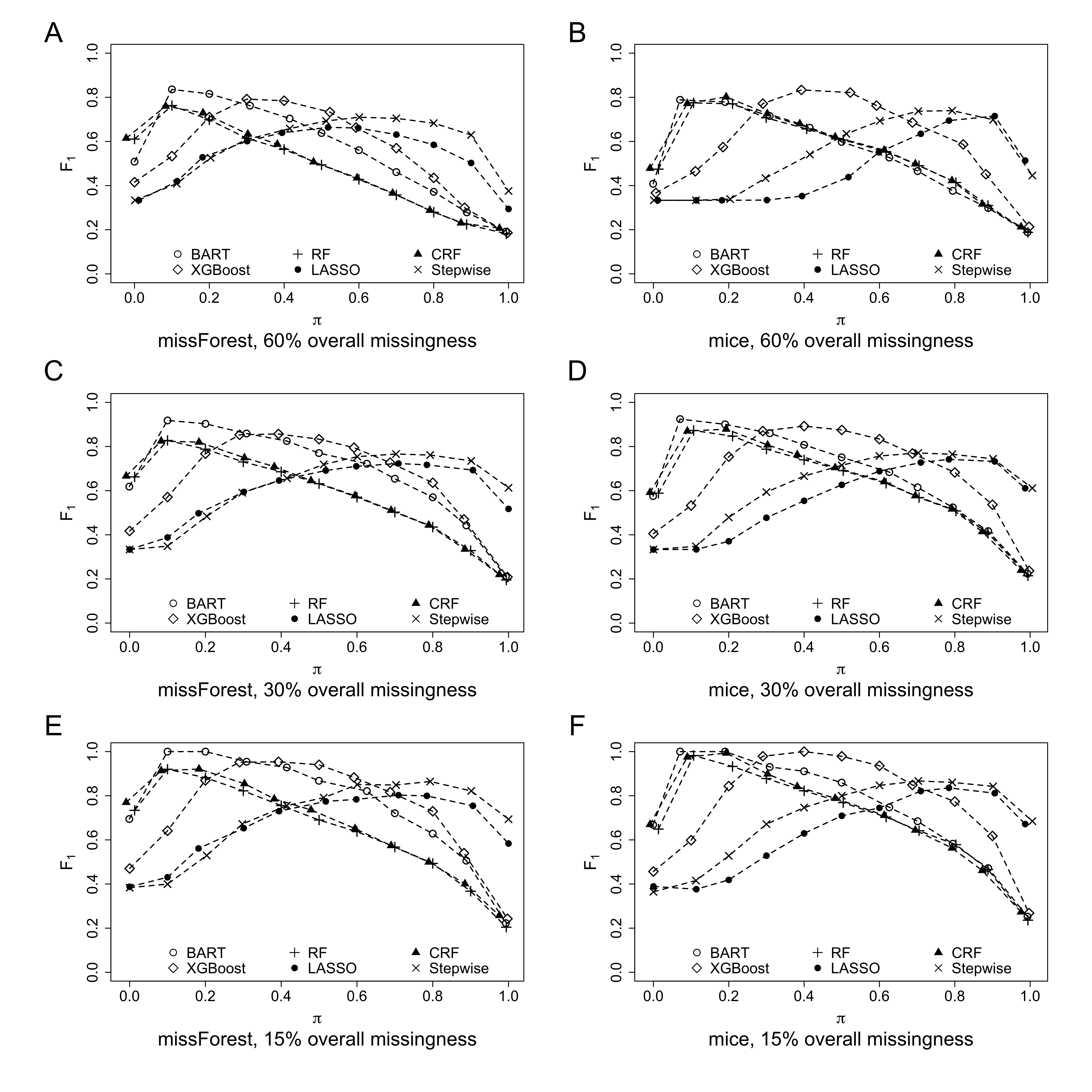}
    \caption{The $F_1$ scores by threshold values of $\pi$ for each of six variable selection methods and for sample size $n=1000$, based on 250 replications.  Imputation was performed on 100 bootstrap samples of each replication data set, using two imputation methods $\code{mice}$ and $\code{missForest}$ for each of three overall missingness proportions, 15\%, 30\% and 60\%.}
    \label{fig:F1_n_1000}
\end{figure}

\begin{figure}[!ht]
    \centering
    \includegraphics[width=1\textwidth]{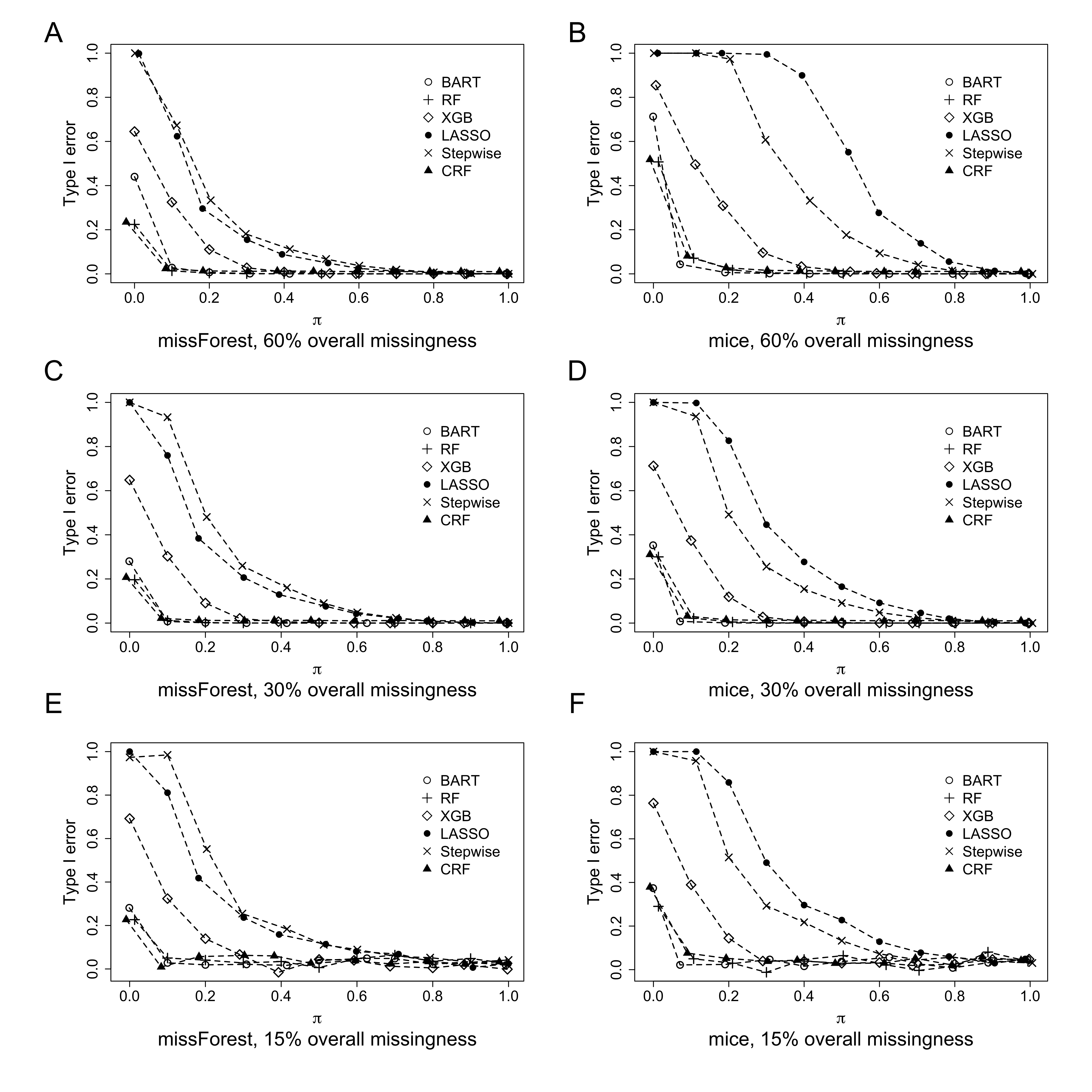}
    \caption{The type I errors by threshold values of $\pi$ for each of six variable selection methods and for sample size $n=1000$, based on 250 replications.  Imputation was performed on 100 bootstrap samples of each replication data set, using two imputation methods $\code{mice}$ and $\code{missForest}$ for each of three overall missingness proportions, 15\%, 30\% and 60\%.}
    \label{fig:TypeI_n_1000}
\end{figure}

\begin{figure}[!ht]
    \centering
    \includegraphics[width=1\textwidth]{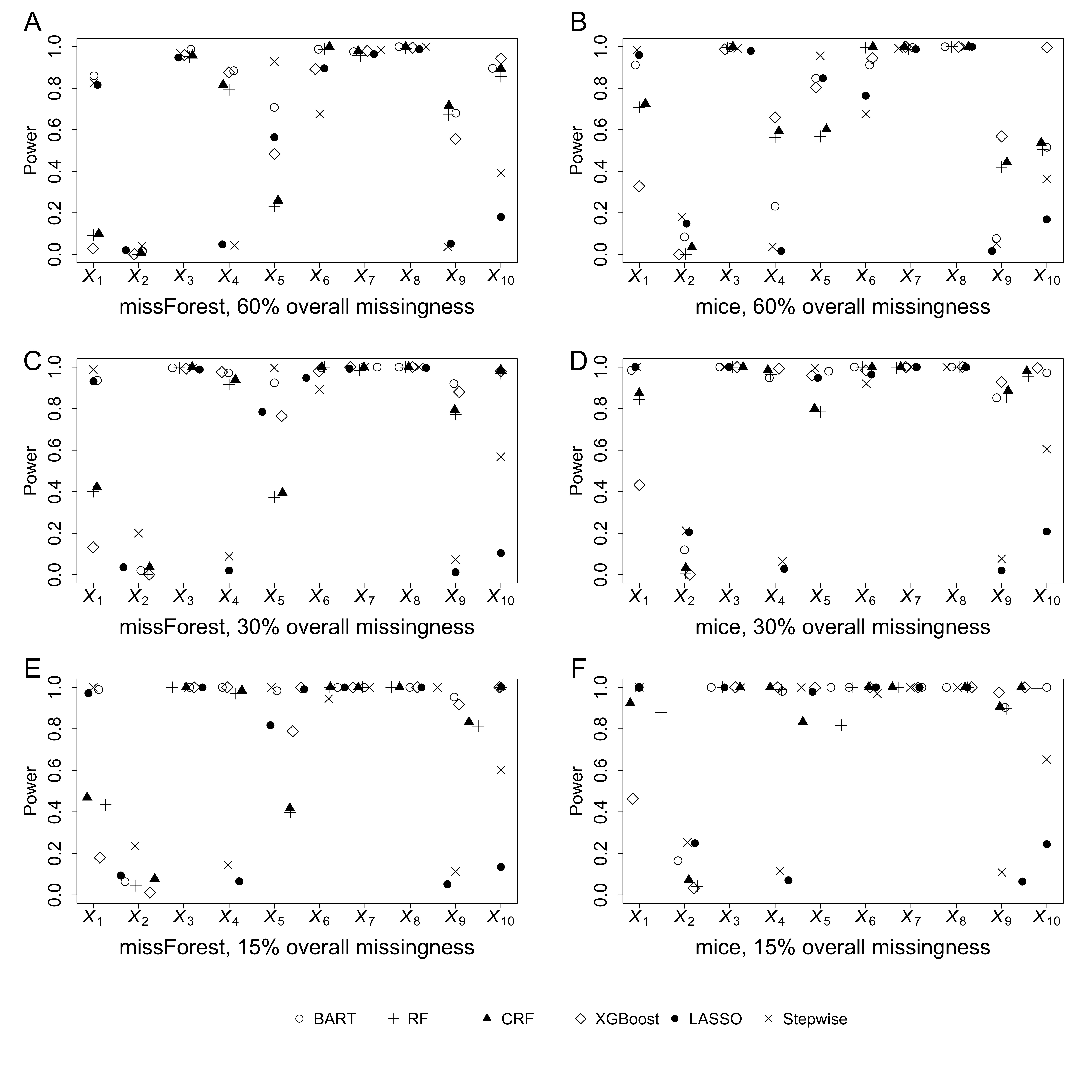}
    \caption{Power of each of six methods for selecting each of 10 useful predictors across 250 replications, when the sample size $n=1000$. Imputation was performed on 100 bootstrap samples of each replication data set, using two imputation methods $\code{mice}$ and $\code{missForest}$ for each of three overall missingness proportions, 15\%, 30\% and 60\%.}
    \label{fig:power_mis_n_1000}
\end{figure}

\begin{figure}[!ht]
    \centering
    \includegraphics[width=1\textwidth]{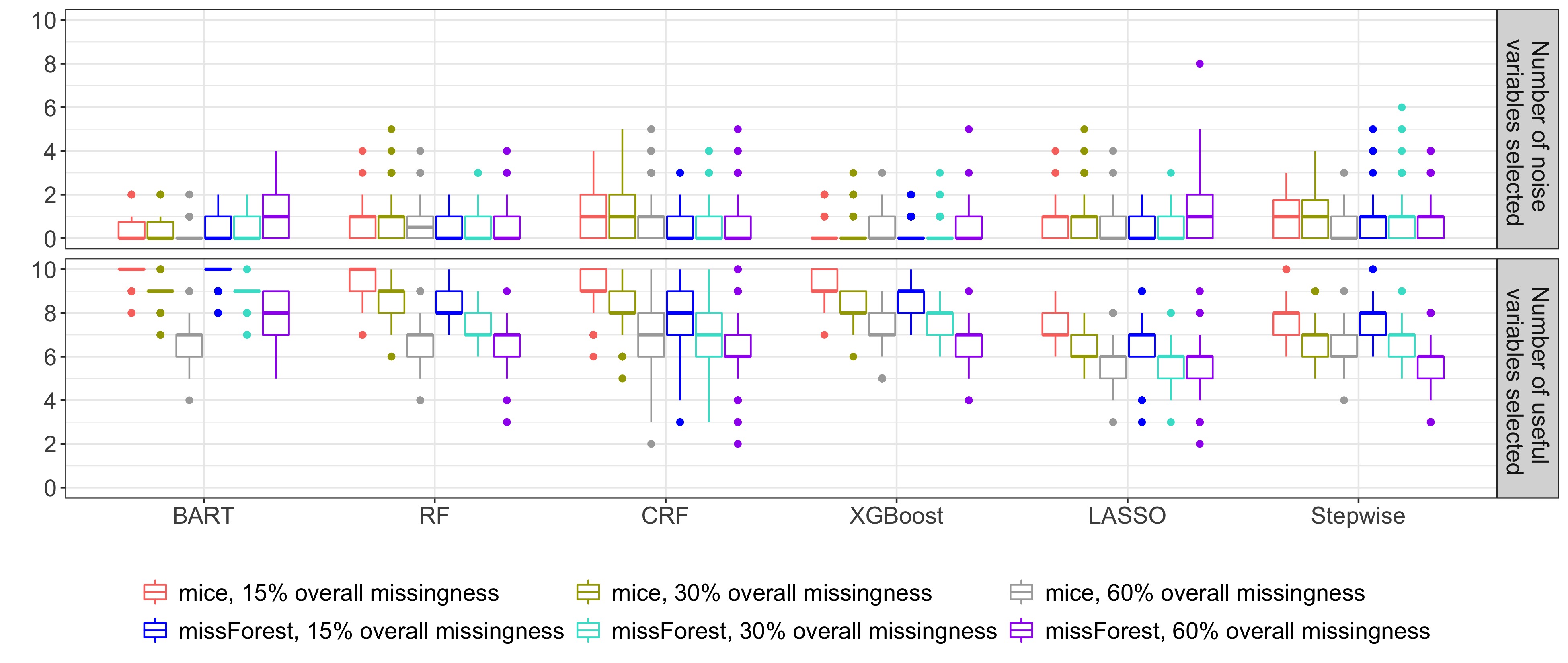}
    \caption{The distribution of the numbers of selected noise predictors and useful predictors for each of six methods in six scenarios and for $n=1000$, across 250 replications. The total number of useful predictors is 10 and the total number of noise predictors is 40.}
    \label{fig:numbers selected_n1000}
\end{figure}

\begin{figure}[!ht]
    \centering
    \includegraphics[width=1\textwidth]{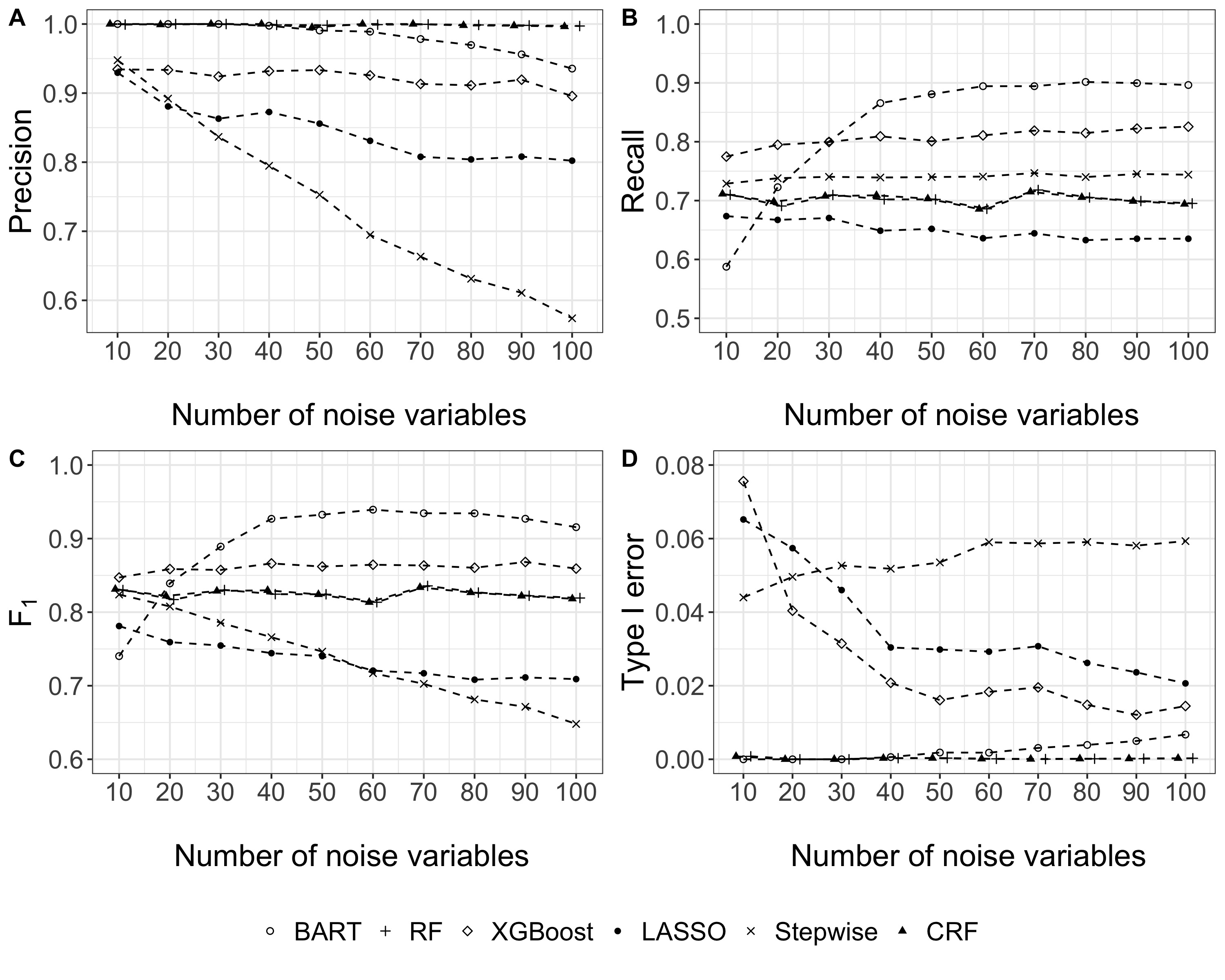}
    \caption{Precision, recall, $F_1$ and type I error of each of six methods on complete data with varying number of noise predictors. The total number of useful predictors is 10, and the total number of noise predictors ranges from 10 to 100. Among the noise predictors,  50\% are  binary variables simulated from Bern(0.5) and 50\% are continuous variables generated from N(0,1). }
    \label{fig:sa_num_noise}
\end{figure}

 With MAR covariates and outcome, our proposed variable selection procedures can recover the performance achieved on complete data. 
 Figure~\ref{fig:precision_n_1000}-\ref{fig:TypeI_n_1000} respectively compare the curves of precision, recall, $F_1$ and type I error over the interval $\pi = (0, 0.1, \ldots, 1)$ for the six methods considered and for $n=1000$.  The variable selection results on incomplete data vary with the threshold value of $\pi$, with the precision increasing and the recall and type I error decreasing over the interval of $\pi$. Overall, BART, CRF and RF have the highest precision and the lowest type I error; backward stepwise selection and lasso have the highest recall but also the highest type I error.  Figure~\ref{fig:F1_n_1000} presents the $F_1$ score, a balance of precision and recall, for each of three missingness proportions and each of two imputation methods. The optimal value of $\pi$ yielding the highest $F_1$ varies by methods, but remains largely consistent across a combination of scenarios of missingness proportions and imputation methods. BART, CRF and RF achieved the best $F_1$ at $\pi = 0.1$ or $\pi = 0.2$,  XGBoost attained the highest $F_1$ with $\pi=0.4$ or $\pi=0.5$, while lasso and backward stepwise selection favored a larger threshold value around $\pi = 0.8$. The performance of all methods deteriorated when the proportions of missing data increased. The nonparametric imputation technique, $\code{missForest}$ did not improve the performance compared to $\code{mice}$, except for BART with 60\% missingness. Figure~\ref{fig:power_mis_n_1000} shows the power of each method for selecting each of the 10 predictors in the $Y$ model. XGBoost has the lowest power among all methods for detecting discrete variables ($X_1$ and $X_2$) across all scenarios, but has superior performance in identifying complex nonadditive ($X_4$ and $X_9$) and nonlinear ($X_{10}$) continuous variables. BART appears to have consistently good performance across all data types, except that in the scenario of 60\% missing data imputed via $\code{mice}$, it has a poor power in detecting $X_4$ and $X_9$.  While having the best performance in selecting discrete variables, lasso and backward stepwise selection still have a low power when the effect size of a discrete variable is moderate ($X_2$). When the sample size reduced to $n=250$, the performance of all methods  deteriorated, but XGBoost appeared to have the smallest drop in the power for detecting the useful predictors, as demonstrated in Web Figure 2. 
 
 Web Figure 3-6 respectively display precision, recall, $F_1$ and type I error for $n=250$. The overall patterns of the four performance metrics remain similar to those for $n=1000$. Although our variable selection procedures on incomplete data can still recover the performance achieved on complete data, with substantially reduced sample size, none of the methods could produce satisfactory performance even with moderate proportion of missing data. Machine learning methods had a larger drop in performance compared to parametric methods.  Among the four tree-based machine learning methods, XGBoost appears to be the least impacted and still outperforms the lasso and backward stepwise selection.

Variable selection results on incomplete data with 15\%, 30\% and 60\% overall missingness are summarized in Table~\ref{tab:simres-mice} for all six methods considered, using the imputation method $\code{mice}$. For each method, we show results for threshold values of $\pi$ leading to the highest $F_1$ and lowest type I error as well as results on the complete cases. Overall, four tree-based machine learning methods produced better performance than lasso and backward stepwise selection, demonstrated by higher $F_1$ and lower type I error. Using the performance on complete data as a benchmark,  all methods can recover the performance benchmark when the missingness proportion is small (15\%) or moderate (30\%) for both sample sizes. With increased proportion of missing data, BART had the largest drop in performance even with a large sample size, and XGBoost appears to be the only method that can achieve the performance benchmark even with a small sample size. Variable selection results using  $\code{missForest}$ as the imputation method are summarized in Web Table 1. The optimal threshold value for $\pi$ tended to be smaller when using $\code{missForest}$, and BART had a better performance when there is a large proportion (60\%) of missing data. Overall,  there is no evident improvement in performance over using $\code{mice}$. For all methods, variable selection via bootstrap imputation yielded better performance than variable selection amongst the complete cases across all scenarios.

Finally, Figure~\ref{fig:numbers selected_n1000} compares the distributions of the numbers of selected noise and useful predictor variables across 250 replications for each of six methods in each of six scenarios defined by missingness proportion and imputation methods and for $n=1000$. Compared to the two parametric methods, lasso and backward stepwise selection, four machine learning methods were able to identify more useful predictors across all scenarios, while generally selecting less noise predictors. Among machine learning methods, the number of selected useful predictors is substantially larger in the case of 15\% missing data than in the case of 60\% missing data, for both imputation methods, $\code{mice}$ and $\code{missForest}$. With the same missingness proportion, there is no apparent ``winner" between the two imputation methods.  

Results of a sensitivity analysis exploring the impact of the number of noise predictors on complete data performance of six variable selection methods are shown in Figure~\ref{fig:sa_num_noise}. The RF and CRF are the only two methods that are comparatively insensitive to the number of noise predictors. For BART, the recall and $F_1$ are highly sensitive to the number of noise predictors; the best performance is achieved when the number of noise predictors is $>30$ (the ratio of useful versus noise predictors $>$ 1/3), and the performance is substantially worse than other methods when the number of noise predictors reduces to 10. XGBoost consistently yields good performance, which is generally insensitive to the number of noise predictors; the Type I error is slightly higher when there are less than 20 noise predictors. Overall, backward stepwise selection and lasso produce lower precision and $F_1$, and higher type I error; and for backward stepwise selection, precision decreases at a fast rate as the number of noise predictors increases. Web Figure 7 shows disaggregated performance of each method for $X_1$-$X_{10}$.  The high sensitivity of BART to lower numbers of noise predictors is manifested in several variables (both discrete and continuous variables). Machine learning methods are less capable of identifying discrete variables, even those with large effect size, but are much more adept at detecting nonlinearity and nonadditivity.   

\section{Case study: The Study of Women's Health Across the Nation}
\label{sec:application}
We analyzed a data set from the Study of Women’s Health Across the Nation (SWAN). The SWAN study was a multicenter, longitudinal study aiming to understand women’s health across the menopause transition. The SWAN data set contains 3302 women of five racial/ethnic groups aged between 42 and 52, who were enrolled in 1996-1997 from seven sites of the US and were followed to 2018 annually. More detail about the SWAN study can be found in Janssen et al.\cite{janssen2008menopause} Despite growing research using the SWAN study data in various areas of women's health, there is a dearth of robust studies identifying key predictors for health outcomes such as metabolic syndrome in the presence of both missing covariates and outcomes. 

Metabolic syndrome is a cluster of conditions that occur together,  representing ``de-tuning'' of metabolic adaptations,  and has been shown to increase the risk of heart disease, stroke and type 2 diabetes \citep{kazlauskaite2020midlife}. 
Prior work \citep{carnethon2004risk,wei2018dietary} has shown that the incidence of metabolic syndrome is associated with various risk factors including blood pressure, triglycerides, waist circumstance, triglycerides, glucose, body mass index, waist to hip circumference ratio, and lipoprotein(a). However, these studies  were not specifically focused on the health of women during their middle years, which was the target population of our study. In addition, it is possible that there exists important risk factors that have not been identified previously.

We sought to identify predictors of 3-year incidence of metabolic syndrome.  Metabolic syndrome is defined as the presence of at least three of the following five symptoms: abdominal obesity, hypertension, hypertriglyceridemia, impaired fasting glucose, and low high-density lipoprotein cholesterol level \citep{kazlauskaite2020midlife}. Our analysis included 2313 women who did not have metabolic syndrome at enrollment. Among the 2313 women, 251 (10.9\%) developed metabolic syndrome within three years of enrollment, 1240 (53.6\%) did not, and the remaining 822 (35.5\%) had missing outcomes. Based on previous literature \citep{han2019dietary,kazlauskaite2020midlife,janssen2008menopause,feng2017low} on risk factors for metabolic syndrome, we selected 60 candidate predictors (29 continuous variables and 31 discrete variables), including  demographics,  daily life behaviour, dietary habits, sleep habits, medications, mental status, menopausal status and related factors, physical measurement, blood measurement, and bone mineral density. A list of 60 variable names and their definitions are displayed in Web Table 2. Only 11 variables are fully observed; the amount of missing data  in the other variables ranges from 0.1\% to 27.1\%. Only 1047 (45.3\%) participants have observed data for all predictors and 763 (33.0\%) for all predictor and outcome variables.

We conducted imputation  on 100 bootstrap samples and implemented six variable selection methods: BART, XGBoost, CRF, RF, backward stepwise selection and lasso. All 60 candidate predictors and the outcome variable were included in the imputation models for both $\code{mice}$ and $\code{missForest}$. The imputation technique and the threshold value of $\pi$, which  gave the best performance for each of the methods in the simulation study under the scenario representative of the SWAN data structure ( $n=1000$ and 60\% overall missingness) , were chosen for each respective method. Table~\ref{tab:case_results} summarizes the variable selection results. BART and XGBoost identified the largest set of predictors (17 and 18); while lasso and backward stepwise selection selected the least predictors (10). Four variables, diastolic blood pressure (DIABP), systolic blood pressure \tb{(SYSBP)}, lipoprotein(a) (LPA) and  triglycerides (TRIGRES), were selected by all of the six methods. Additional four variables  were selected by all four machine learning methods, body mass index \tb{(BMI)}, tissue plasminogen activator (TPA), waist circumference (WAIST) and waist to hip circumference ratio (WHRATIO), among which, TPA  and WHRATIO were not identified by the two parametric methods. The backward stepwise selection approach selected more variables that were not chosen by any of four machine learning methods than the lasso. A full list of names and definitions for the 60 candidate predictor variables can be found in Web Table 2.

\begin{figure}[!ht]
    \centering
    \includegraphics[width=0.5\textwidth]{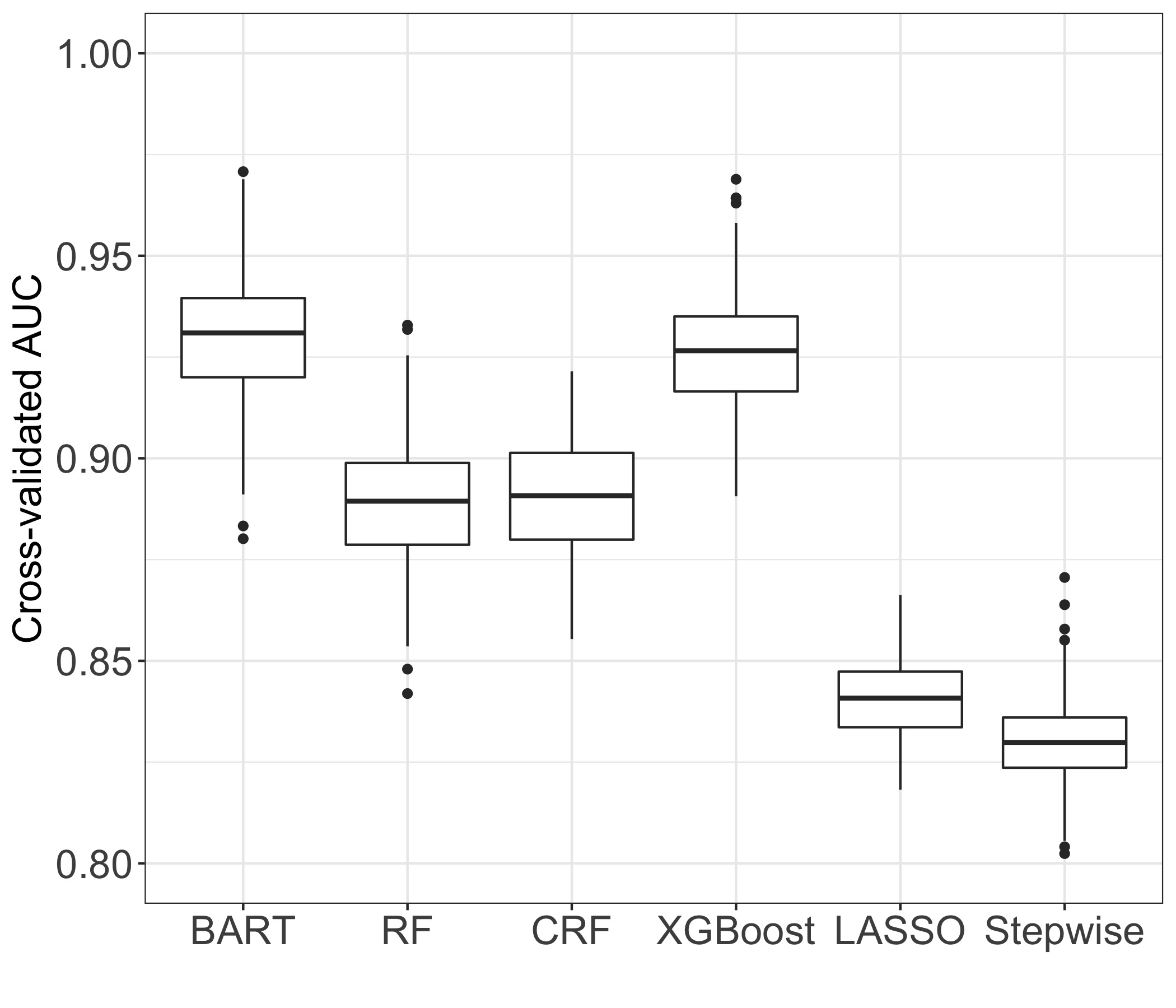}
    \caption{The distribution of 5-fold cross-validated area under the curve (AUC) among 100 bootstrap samples of the SWAN dataset with imputation performed for each of six methods.}
    \label{fig:SWAN_AUC}
\end{figure}

As an anonymous reviewer pointed out,  in the situation where it remains unclear  which method is able to select the most relevant predictors, one may use the cross-validated error to help distinguish between the methods. We evaluated the prediction performance, based on area under the curve (AUC), Cohen's Kappa statistic and misclassification error, of the six models that regress the outcome variable on their respectively selected predictor variables.  Figure~\ref{fig:SWAN_AUC} shows the distribution of 5-fold cross-validated AUC among 100 bootstrap samples with imputation performed via the best technique suggested in simulations for each of six methods. BART boasts the highest AUC, closely followed by XGBoost; the lasso and backward stepwise selection have the lowest AUC. Displayed in Web Figure 8-9 are the distributions of 5-fold cross-validated Cohen's Kappa  and misclassification error, which suggest the same performance rankings of the methods. In light of these results, variables selected by BART may be an important addition to the literature. Among the 17 variables selected by BART, some have been less commonly identified as risk factors for metabolic syndrome in the literature, but selection of these variables may be supported by domain knowledge. For example,  levels of TPA antigen and low apolipoprotein A-1 (APOARES) were found to be associated with insulin resistance, which was involved in the pathogensis of impaired fasting glucose \citep{rao2004impaired, feng2017low}.

\begin{table}[htbp]
    \centering
    \caption{Variable selection results by each of six methods, with the best imputation method and threshold value of $\pi$ suggested in simulations. BART was used with $\code{missForest}$ and $\pi=0.1$, CRF and RF with $\code{mice}$ and $\pi=0.2$, XGBoost with $\code{mice}$ and $\pi=0.5$, lasso with $\code{mice}$ and $\pi=0.9$, Stepwise with $\code{mice}$ and $\pi=0.8$. Definitions of the variable names appear in Web Table 2.} 
    \bgroup
\def\arraystretch{1.25} 
    \begin{tabular}{p{2.0cm} p{1.5cm} p{12.5cm}}
    \toprule
        Methods &\# selected variables &Selected variables  \\
        \midrule
         BART & 17 & APOARES, BMI, BP, DIABP, DTTLIN, GLUCOSE, LPA, RACE, RESTLES, SHBG, SPBMD, SYSBP, T, TPA, TRIGRES, WAIST, WHRATIO\\
           XGBoost  & 18 & APOARES, BMI, DIABP, EDUCATION, E2AVE, GLUCOSE, HPBMD, INSULIN, LMPDAY, LPA, PAI1, SHBG, SPBMD, SYSBP, TPA, TRIGRES, WAIST, WHRATIO \\ 
      CRF  & 11 & BMI, DIABP, EDUCATION, INSULIN, LPA, PAI1, SYSBP, TPA, TRIGRES, WAIST, WHRATIO \\       
         RF  & 12 & BMI, CRP, DIABP, EDUCATION, INSULIN, LPA, PAI1, SYSBP, TPA, TRIGRES, WAIST, WHRATIO \\
         Stepwise  & 10 & BMI,  DIABP, EDUCATION, LPA, NOINSURE, PHYSWORK, RESTLES, SPORTS, SYSBP, TRIGRES \\          
         lasso  & 10 & ALLCALC, APOARES, DIABP, DTTSWET, E2AVE, GLUCOSE, LPA, SYSBP, TRIGRES, WAIST \\ 
         \bottomrule
    \end{tabular}
  \egroup  
    \label{tab:case_results}
\end{table}

\section{Discussion} \label{sec:discussion}
We investigate a general variable selection approach when there are missing data in both covariates and outcomes. This approach exploits the flexibility of machine learning modeling techniques and bootstrap imputation, which is amenable to nonparametric methods in which the effect sizes of predictor variables are not naturally defined as in parametric models. Our numeric results show that the proposed variable selection procedure based on three machine learning methods: BART, RF and XGBoost achieves good practical operating characteristics. When the sample size is sufficiently large (e.g., $n=1000$), even with a large amount of missing data, the variable selection procedure can recover the performance achieved on complete data. The proposed approach can be readily applied to a variable selection problem with a general missing data pattern. 

Several considerations must be taken into account in choosing an ``optimal'' variable selection method. First, an investigator needs to decide the goal of variable selection and choose the corresponding performance metrics: whether it is to identify as many useful predictors as possible (precision), or to avoid selecting irrelevant predictors (recall), or to achieve the best balance between these two goals ($F_1$ and type I error). Second, while BART has the best performance across various scenarios in our simulation study, our exploratory sensitivity analysis on complete data shows that BART is sensitive to the ratio of useful versus noise predictors. When the ratio is relatively large (e.g., 1:1), BART has a poor recall and $F_1$. As the ratio decreases to $<1/3$, BART achieves the best balance between selecting useful predictors and avoiding irrelevant predictors. On the other hand, XGBoost shows good performance across various scenarios and is generally insensitive to the amount of noise information, but has a low power of detecting discrete predictors. RF is highly capable of avoiding selecting unimportant predictors but is less capable of uncovering the full set of important predictors. In practice, it may be useful to run several variable selection approaches and if the chosen variable selection performance metrics are similar across methods, then a cross-validated error estimation of the model with selected variables may help further distinguish between methods, as demonstrated in our case study.  Third, while adept at detecting nonlinear and nonadditive variables, machine learning methods have lower power of detecting discrete variables, even those with relatively strong effects, compared to parametric models such as lasso and backward stepwise selection. As a result, machine learning methods may be less favored if a majority of candidate predictors are discrete. 

One limitation of our bootstrap imputation-based variable selection procedure  is the computational cost of running machine learning models on a large data set. However, it should be noted that the bootstrap resampling can be computed in parallel on multiple cores when such resources are available. A challenging but important avenue for future research is developing inference-based variable selection methods using the variable importance measure provided by machine learning models.  An immediate next step could be to extend the BART-based approach by leveraging the Bayesian posterior samples of the variable inclusion proportions. When using multiple imputation for missing data in conjunction with BART models, we can use Rubin's rule \citep{wood2008should, hu2019causal} or pooled posterior samples arising from the multiple datasets \citep{hu2020flexible} to combine inferences for variable inclusion proportions, which can be further used for variable selection. It is less straightforward for frequentist machine learning methods.   
Ishwaran and Lu \cite{ishwaran2019standard} proposed a subsampling approach to estimate the variance and construct confidence intervals of RF's variable importance scores. However, in the presence of missing data, the reduced sample size in a random subsample should have an adverse effect on imputation as it depends on the observed data \citep{long2015variable}. One possible strategy is to derive a nonparametric extension of the usual ANOVA-derived measure of variable importance in parametric models. Particularly, Williamson et al. \citep{williamson2021nonparametric} discussed a generalization of
the ANOVA variable importance measure, $\Psi_s(P) = \int \{\mu_P(x) - \mu_{P,s}(x)\}^2dP(x) /var_P(Y)$, where 
$\mu_P(x)$ and $\mu_{P,s}(x)$ are respectively the conditional mean of outcome $Y$ given the full set and a subset of covariates, under the data-generating mechanism $P$. Further research in adapting this measure into the settings where missing data are present could be a worthwhile contribution.

\begin{dci}
The authors declare that they have no conflict of interest.
\end{dci}

\begin{funding}
 This work was supported in part  by award ME\_2017C3\_9041 from the Patient-Centered Outcomes Research Institute, and by grants R21CA245855 and P30CA196521-01 from the National Cancer Institute. 
\end{funding}


\newpage
\bibliographystyle{SageV} 
\bibliography{references}



\end{document}